\documentclass[aps,prl,showpacs,float]{revtex4-1}
\usepackage{graphicx}
\begin{document}
\title{NUMERICAL SOLUTIONS OF THE QUANTUM HAMILTON-JACOBI EQUATION AND WKB-LIKE REPRESENTATIONS FOR ONE-DIMENSIONAL WAVEFUNCTIONS}
\author{Mario Fusco Girard}
\affiliation{Department of Physics ''E.R. Caianiello'',\\University of Salerno \\and \\Gruppo Collegato INFN di Salerno,\\Via Giovanni Paolo II, 84084 Fisciano (SA), Italy}

\begin{abstract}
 By means of  numerical solutions of the quantum Hamilton Jacobi equation, a  general WKB-like representation for one-dimensional wavefunctions is obtained. This representation is unique in the classically forbidden regions, while in the allowed one, each wave function corresponds to a one-parameter family of solutions of the QHJE. The method has been applied to various systems, with different energies and initial conditions. In all investigated cases, the wavefunctions so obtained accurately reproduce the solutions of the Schroedinger equation, analytically or numerically computed by other ways. Some results for harmonic oscillator and radial Coulomb motion are presented.
\end{abstract}

\maketitle

\section{}
Due to its key role for the classical limit of the Schroedinger Equation (SE) [1] and in the WKB approximation method [2], the Quantum Hamilton-Jacobi Equation (QHJE)  attracted the attention of researches soon after  the birth of Quantum Mechanics (QM). The interest was renewed in the 80's by a series of papers by Leacock and Padgett [3,4], and by Floyd [5-9]. In Ref. [3,4] a formulation of QM based on the QHJE was presented, which allows to find the energy levels for one-dimensional potentials, without solving the SE. In [5-9] the QHJE was investigated in the framework of a hidden variables theory, and the relationships with the Bohm's  formulation of QM [10-11] were discussed. The approach of Refs. [3,4] was extended, showing that also the eigenfunctions for a  wide class of potentials  can be found [12-18]. A QHJE for the generating function of a canonical transformation of phase-space path integral was proposed [19]. An operatorial QHJE was studied [20], and  extended to relativistic case [21]. The relationships between  QHJE and Supersymmetric QM were discussed [22-24]. Analytical solutions of QHJE, in terms of known solutions of SE were found [25-27]. Quantum trajectories derived from the QHJE were applied to the study of atom diffraction[28]. An accurate computational method  was  developed for bound state wave functions [29] and for scattering problems [30]. The QHJE was applied to the study of the quantum particles' trajectories [31-40]. A proposal to reconcile the semiclassical and Bohmian mechanics was presented and developed [41-43]. A geometric phase within the QHJ formalism was introduced [44,45]. Quantum interference was investigated [46,47]. Wave front-ray synthesis for solving the multidimensional QHJE was presented [48]. Finally, the QHJE was also derived, within a geometric-differential approach, from an equivalence principle [49-50].

The aim of  this Letter is to present some results of a numerical investigation of the QHJE for one-dimensional or separable motion.  As the construction of  Hamiltonian wavefronts, some properties of the classical action, its role in the formation  of  semiclassical wavefunctions for non separable potentials have been investigated in some previous paper [51-55], the present work originated firstly  to compare the quantum action with the corresponding classical one.

As in the usual WKB method, the starting point is the search for exponential-type solutions
\begin{equation}
\psi(x)=e^{{i\over \hbar} W(x)}
\end{equation}			 
of the time-independent SE for one dimensional motion in a potential $V(x)$:
\begin{equation}
-{\hbar ^2\over 2m} {d^2\psi\over dx^2}=\left[E-V(x)\right]\psi\ .
\end{equation}    
Substitution of Eq. (1) in (2) gives the one-dimensional time-independent QHJE:
\begin{equation}
{1\over 2m}\left({dW\over dx}\right)^2 - {i\hbar \over 2m}{d^2W\over dx^2}=E - V(x)\ .
\end{equation}  

Its solution  $W(x)$  is the quantum characteristic function or (reduced) action of the particle.
For $\hbar = 0$,  Eq. (3) reduces to the classical Hamilton-Jacobi equation  [56] for the characteristic function  $W0(x)$
\begin{equation}
{1\over 2m}\left({dW0\over dx}\right)^2 = E - V(x)\ ,
\end{equation} 
whose solution is the elementary integral on the particle's classical momentum $p(x)$
\begin{equation}
W0(x) = \int p(x) dx = \pm\int \sqrt{2m \left({E- V(x)}\right)} dx\ .
\end{equation} 

Let us suppose for simplicity that the potential  has only two turning points, $x1$ and $x2$, with $E>V(x)$  for $x1<x<x2$. Three regions are so  defined: I): $x<x1$;   II):      $x1\le x\le x2$     ;      III):       $x2<x$.

In the WKB method, the phase $W(x)$ in Eq. (1) is formally  written as a power series of  $(\hbar / i)$, then inserted into the QHJE; by retaining as usual only the first two terms, the WKB approximate representation of the wavefunction is obtained, which in region II, with the choice $W0(x1)=0$, is:
\begin{equation}
\psi_{WKB,II}(x) ={A\over \sqrt{|p(x)|}}\sin\left[{W0(x)\over \hbar}+{\pi\over 4}\right] +O(\hbar ^2)\ ,
\end{equation}        
with $A=\rm const$.
Two analogous expressions, but with exponential functions of the appropriate $W0_{I,III}(x)$ instead of the trigonometric function, hold for the other regions.
The present approach is similar to the WKB one, but  we are instead looking for exact solutions of the QHJE; in regions I and III the wavefunctions are exponentially damped, so we search for  purely imaginary solutions  $W _{I,III}=iY_{I,III}$; Eq. (3) then reduces to:
\begin{equation}
-Y'^2+\hbar Y''= 2m(E-V(x))\ ,
\end{equation}        
(apices denoting derivatives with respect to $x$). Therefore, the wavefunctions  in regions I and III  have the exponential type representation:
\begin{equation}
\psi_{I,III}(x)=B_{I,III}e^{-Y_{I,III'}(x)/\hbar}\ ,
\end{equation}     
with $B_i=\rm const$.

In regions II,  looking for oscillating behavior of wavefunctions, we write instead the phase in full complex form
\begin{equation}
W(x) =X(x)+i Y(x)\ .
\end{equation}        
By substituting this into Eq. (3), and by separately equating the real and the imaginary parts of the two sides we get
\begin{equation}
X'^2 - Y'^2 + \hbar Y'' =2m\left({E-V(x)}\right)\ ,
\end{equation}     
\begin{equation}
X'Y' -{1\over 2}\hbar X'' = 0\ .
\end{equation}   

Equation (11) is immediately integrated to
\begin{equation}
Y(x) =\hbar\log \left[\sqrt{|X'(x)|}\right] + {\rm const}\ .
\end{equation}   
Substituting this into Eq. (10) gives a non linear third order equation  for the real part X of the quantum action W [57]
\begin{equation}
X'^2 - {3\over 4}\hbar ^2 {X''^2 \over X'^2} + {1\over 2} \hbar ^2 {X''' \over X'} = 2m\left(E -V(x) \right)\ .
\end{equation}   
Once a suitable solution $X(x)$  of this equation in region II is known,  the imaginary part Y  is obtained from Eq. (12). This gives the quantum action $W=X+iY$  in the classical region, which, inserted in Eq. (1), provides a fundamental solution of the SE. By combining this solution and its conjugate, we write the wavefunction for the region II in the WKB  form:
\begin{equation}
\psi_{II}(x)=\frac{B_{II}}{\sqrt{|X'(x)|}}\sin\left[X(x)+{\pi\over 4}\right]\ .
\end{equation} 		     

Eqs. (8) and  (14), with appropriate functions $X(x)$ and $Y_{I,III}(x)$ give an exact representation of the wavefunction. Differently from the WKB one, in the quantum case this representation is  however not unique (see below).
Comparison between Eqs. (6) and (14) shows that the function X(x) corresponds to the classical characteristic function W0(x), and its derivative $X'(x)$ is analogous to the classical momentum $p(x)$.

While the Riccati  Eq. (7) for some potential can be analytically integrated, this task for Eq. (13) is in general much more difficult; the numerical  integration of these equations, however, presents no particular problems. The computation has been done for various potentials, with different energies and initial conditions. In any case, the wavefunctions computed by the present method, very well agree with the analytical or numerical corresponding solutions of the SE (with standard numerical procedures the error is less than $10^{-7}$).  Some results, for the harmonic oscillator and for the radial motion in a Coulomb potential are presented in the figures. A more extended report will be given elsewhere [58].

To start the numerical integration, the energy E and the initial value $X(x1)$ have to be fixed: this latter is largely arbitrary because a constant term can be added to the function $X(x)$. For analogy with the classical case, in the figures we chose $X(x1)=W0(x1)=0$, but different choices are possible, as discussed below. The other needed quantities can then be obtained from the vanishing of the wavefunctions at $x= \pm \infty$, the continuity of these and their first derivative at the turning points, and the normalization condition. This is equivalent to use the values of the solution of the SE and its derivative at the turning points, analytically known or numerically computed by other ways. In any case, it is easy to see that while the functions $Y_{I,III}(x)$ giving the representation (8) in the regions I and III, are in correspondence one-to-one with the wavefunction,  this is not true in the classical region for the function $X(x)$. Indeed, when the known values $\psi(x1)$, and $\psi'(x1)$ are inserted in Eq.( 14) and its derivative, the quantity $X(x1)$ can be arbitrarly chosen in the interval of values, and this in correspondence fixes  the initial  values for $X'(x1)$ and $X''(x1)$. Therefore, each choice for $X(x1)$ corresponds to a different solution of Eq. (13) at fixed energy; the numerical integration however confirms that all these solutions give  the same combination in Eq. (14), i.e. the wavefunction itself  is  left unchanged. In summary, each eigenfunction corresponds to a one-parameter family of solutions of the QHJE in the classical region. This is due the fact that, while SE is a second order equation, Eq. (13) is a third-order one.

Figures 1-3  refer to the harmonic oscillator $V(x)= {1\over 2} x^2$, while Fig. 4 is for radial motion in a  Coulomb potential $1/r$. In the computations, the values $\hbar =1$ for the Planck's constant, and  $m=1$ for the particle's mass were used. In Fig. 1 the real part $X(x)$ of the quantum characteristic function $W(x)$ for the state $n=8$ of the oscillator in the region II is reported, together with the corresponding classical quantity $W0(x)$.  Fig. 2 shows the derivative $X'(x)$ and the classical momentum $p(x)=W0'(x)$. In Fig. 3 are plotted the quantities $1/\sqrt{|X'(x)|}$, $\sin[X(x)+ \pi /4]$   and finally their product, which gives according to Eq. (14) the  eigenfunction in the classical region. For simplicity the left and right exponential tails are non reported.

 \begin{figure}[h]
 \includegraphics[scale=0.8]{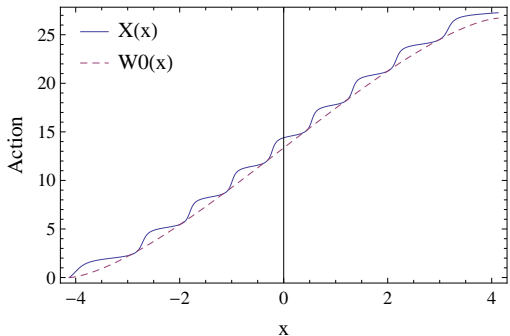}%
 \caption{\label{}The real part $X(x)$ of the quantum action (continuous line) and the classical characteristic function $W0(x)$ (dashed line) for the $n=8$ state of the harmonic oscillator.}
 \end{figure}
 \begin{figure}[h]
 \includegraphics[scale=0.8]{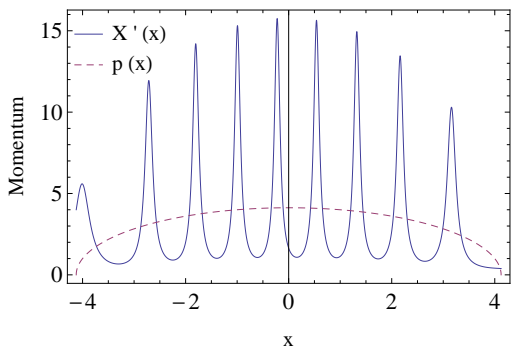}%
 \caption{\label{}The derivative $X'(x)$ (continuous line) and the classical momentum $p(x)$ (dashed line) for the $n=8$ state of the harmonic oscillator.}
 \end{figure}
 \begin{figure}[h]
 \includegraphics[scale=0.8]{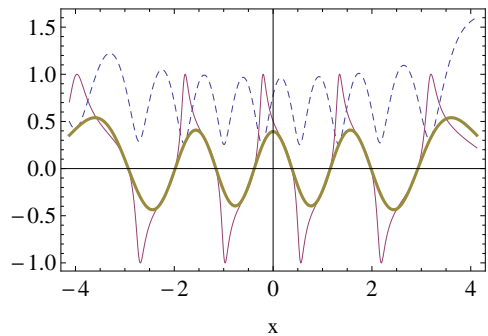}%
 \caption{\label{}The quantities  $1/\sqrt{|X'(x)|}$ (dashed line), $\sin[X(x)+ \pi /4]$ (thin line) and their product (thick line), which according to Eq. (14) represents the wave function in the classical region, for the n= 8 state of the harmonic oscillator.}
 \end{figure}
This wavefunction accurately reproduces the well known $n=8$ state of the oscillator.
As seen from the figures, $X(x)$ is a staircase increasing function which follows the profile of the classical action $W0(x)$, each "step" corresponding to a peak in the derivative $X'(x)$. The phase of the wavefunction in Eq. (14) starts  from  $\pi /4$  at the left turning point and increases with  x.  Each passage of the phase through a multiple of $\pi$ gives a node of the wavefunction, so that the positions of these latter are strictly controlled  by the values of X(x). In general, with the choice X(x1)=0,  the number of steps and peaks is equal to the number $n+1$, where $n$ is the number of nodes of the wavefunction: in effect, the first step and peak at left in the figures (1) and (2), respectively, are due to our choice of the initial value for the phase in Eq. (14) and can be smoothed or completely eliminated through different choices. Finally,  Fig. 4 is for the radial motion with  $n=3$, $l=1$ in a Coulomb potential and shows the effectiveness of the method also for  higher-dimensional separable potentials. In the integration the value $X(r1)=0$ at the left turning point r1 was maintained,  but a different value for the initial  phase in Eq. (14) was chosen, i.e. $\phi =0.01$ instead of $ \phi= \pi /4$. This gives a smoother staircase behaviour for the radial action $X(r)$, and correspondingly smaller peaks in the derivative $X'(r)$. In Fig. 4, the upper box presents the solutions of the QHJE in the three regions I, II and III; in the lower one, the resulting  wavefunction is plotted, which also in this case accurately reproduces the corresponding  solution of the SE.
 \begin{figure}[h]
 \includegraphics[scale=0.8]{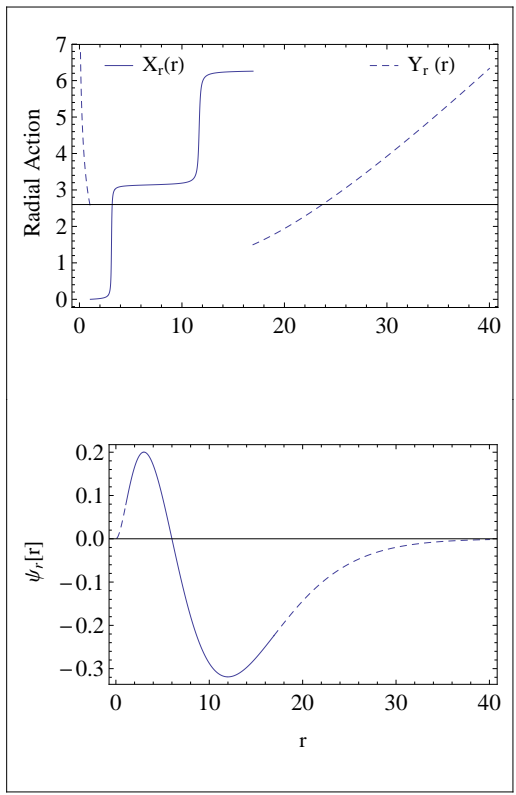}%
 \caption{\label{}Upper box: the solutions $Y_I(x)$, (dashed line, left), $X(x)$ (continuous line) and $Y_{III}(x)$ (dashed line, right) of the QHJE for the radial motion with  $n=3$, $l=1$ in a Coulomb potential. Lower box: the corresponding radial part of the eigenfunction, constructed from the solutions of the QHJE plotted in the upper box, by means of Eq. 8 ( regions I and III, dotted lines) and by means of Eq. (14) for the region II (continuous line).}
 \end{figure}
The general exact quantization condition for one-dimensional systems  has been formulated in Refs. [3,4]. In the present approach, the (exact) quantization condition, as the WKB (approximate) one,  follows from the fact that, for a generic value of the energy E, it is  not possible to smoothly join the solution (14), constructed for region II starting from the first turning point, with the corresponding $\psi_{III}(x)$, for the region III, given by Eq. (8). In terms of the values of the function $X(x)$, this condition can be put in various equivalent forms, according to the potential V(x), but differently from the WKB case, it in general involves not only the values $X(x1)$, $X(x2)$,  but also the derivatives $X'(x1)$, $X'(x2)$.
A more extended discussion of the quantization condition will be given elsewhere.

\end{document}